\documentclass[sn-basic,authoryear]{sn-jnl}

\usepackage{graphicx}%
\usepackage{verbatim}
\usepackage{color}
\usepackage{multirow}%
\usepackage{amsmath,amssymb,amsfonts}%
\usepackage{amsthm}%
\usepackage{mathrsfs}%
\usepackage[title]{appendix}%
\usepackage{xcolor}%
\usepackage{textcomp}%
\usepackage{manyfoot}%
\usepackage{booktabs}%
\usepackage{algorithm}%
\usepackage{algorithmicx}%
\usepackage{algpseudocode}%
\usepackage{listings}%

\newcommand{\R}{\mathbb{R}} 
\newcommand{\Q}{\mathbb{Q}} 
\newcommand{\N}{\mathbb{N}} 
\newcommand{\PP}{\mathbb{P}}
\newcommand{\EE}{\mathbb{E}}
\newcommand{\E}{\mathbb{E}}

\newtheorem{theorem}{Theorem}
\newtheorem{proposition}[theorem]{Proposition}%
\newtheorem{example}{Example}%
\newtheorem{remark}{Remark}%
\newtheorem{problem}{Problem}

\newtheorem{definition}{Definition}%

\begin{document}

\title[MDP with Risk-Sensitive Criteria: An Overview]{Markov Decision Processes with Risk-Sensitive Criteria: An Overview}


\author*[1]{\fnm{Nicole} \sur{B\"auerle}}\email{nicole.baeuerle@kit.edu}

\author[2]{\fnm{Anna} \sur{Ja\'skiewicz}}\email{anna.jaskiewicz@pwr.edu.pl}

\affil*[1]{\orgdiv{Department of Mathematics}, \orgname{
Karlsruhe Institute of Technology (KIT)}, \orgaddress{
\city{Karlsruhe}, \postcode{76131}, \country{Germany}}}

\affil[2]{\orgdiv{Faculty of Pure and Applied Mathematics}, \orgname{Wroc{\l}aw University of Science and Technology}, \orgaddress{ \city{Wroc{\l}aw},  \country{Poland}}}


\abstract{The paper provides an overview of the theory and applications of risk-sensitive Markov decision processes. The term 'risk-sensitive' refers here to the use of the  Optimized Certainty Equivalent as a means to measure expectation and risk. This comprises the well-known entropic risk measure and Conditional Value-at-Risk. We restrict our considerations to stationary problems with an infinite time horizon. Conditions are given under which optimal policies exist and  solution procedures are  explained. We present both the theory when the  Optimized Certainty Equivalent is applied recursively as well as the case where it is applied to the cumulated reward. Discounted as well as non-discounted models are reviewed. }

\keywords{Markov decision process, Risk-sensitive decision,  Optimized Certainty Equivalent, Optimal policy}



\maketitle

\section{Introduction}\label{sec1}
The theory of Markov decision processes (MDPs) deals with 
stochastic, dynamic optimization problems. In the classical 
situation, the aim is to maximize an expected cumulated or 
averaged reward of a system. Since the first formulations by 
Richard Bellman in the 1950s, the theory has developed 
tremendously. In particular, one branch of literature is 
devoted to extending this theory beyond the simple expectation, 
since there is an evidence from various fields that the 
expectation should be replaced by some criterion which allows to 
model risk-sensitivity of the decision maker. This evidence 
comes from disciplines like psychology, economics and biology. 
For instance, \cite{braun2011risk}  reviewed  evidence for risk-sensitivity in motor control tasks. 

From a mathematical point of view, the decision problem gets of 
course more complicated when risk-sensitivity is taken into 
account. Loosely speaking, risk-sensitivity weights the 
possible fluctuations around the mean. A simple way to deal 
with this is to consider a weighted criterion of the expectation 
and the variance of a random income, i.e. to include the second moment into the decision. 
This has for example been propagated in \cite{markowitz1952}. 
Naturally, one can  generalize this idea to higher moments. One of the ways is to use
an exponential function which plays  
a prominent role in risk-sensitive MDPs.
Then, all moments of a random payoff are taken into account if we consider the expectation of an  
exponential function of this random  payoff. This fact can be seen  via the
Taylor series expansion of the exponential function around 0. 
To be more precise let us consider  for example the following expression
$$ J(x,\pi)=-\frac1\gamma\ln \EE_x^\pi\left[  \exp\left(-\gamma  \sum_{k=0}^\infty \beta^k r(X_k,A_k)\right) \right]$$
where $(X_k,A_k)_k$ is a controlled state-action process, $r$ is a one-stage reward function, $\beta$ a discount factor, $\gamma\not=0$ is a risk-sensitivity 
parameter and the transition law is determined by a policy $\pi$. The initial state is $X_0=x$.  A target function like this has first been studied in \cite{howard1972risk}. Indeed, for small $\gamma$ this is approximately equal to 
$$ J(x,\pi) \approx \EE_x^\pi\left[ \sum_{k=0}^\infty \beta^k r(X_k,A_k)\right] - \frac\gamma2 Var_x^\pi\left(\sum_{k=0}^\infty \beta^k r(X_k,A_k)\right).$$
However,  from a mathematical point of view it is more 
tractable than the variance. From the approximation it is 
also obvious that $\gamma >0$ models a risk-averse decision 
maker (since then the variance is subtracted), whereas 
$\gamma <0$ corresponds to a risk-loving decision maker. The 
preceding target function is a special case of the  situation we 
consider here in this paper. It can also be interpreted as a 
Certainty Equivalent of the exponential utility function. 
This point of view can then be generalized to Optimized 
Certainty Equivalents which we consider in this survey. 

The aim of this paper is to provide an overview of the 
ideas, concepts and literature in this area. We will also 
discuss the situation where the Optimized Certainty 
Equivalent is applied to the single-stage rewards in a 
recursive way. However, we will stay within the setting 
where optimal policies are stationary in a certain sense and can be computed 
from optimality equations, thus naturally avoiding time-inconsistency issues. Our point of view is mainly from the 
economics and operations research perspective. We do not 
consider problems with a finite time horizon, nor do we 
treat problems in continuous time. For this direction the 
reader is referred to 
the recent survey by \cite{biswas2023ergodic}.

The outline of our survey is the following. 
In the next section 
we explain and discuss our main building block for the 
target function: the Optimized Certainty Equivalent. The 
Optimized Certainty Equivalents have been introduced by 
\cite{ben2007old} and provide a useful generalization of  
Certainty Equivalents. They comprise important cases like 
the entropic risk measure and the 
Conditional Value-at-Risk and 
are still tractable from a mathematical point of view. In 
Section \ref{sec:MDP} we introduce the  theory of Markov 
decision processes. We restrict our attention to stationary problems (i.e.\ the model data do not depend on the time point) with an infinite time horizon. Conditions are given under which optimal policies exist and a solution procedure is explained. Section \ref{s:mdp_rec} presents the theory when the  Optimized Certainty Equivalent is applied recursively. Some generalizations and related problems are discussed at the end. Afterwards, Section \ref{sec:outside} treats the situation when the  Optimized Certainty Equivalent is applied to the cumulated reward. Here the presented solution technique is via an extension of the state space. Finally in Section \ref{sec:other} we provide an overview on the risk-sensitive average cost case. Section \ref{sec:appl} summarizes some typical applications of the presented theory. The appendix contains two proofs.

{\bf Notation.} As usual, the symbol $\N$ denotes the set of positive integers and $\N_0=\N\cup\{0\}.$ By $\R$ we denote the set of all real numbers. We use the following 
abbreviations: {\it w.r.t.} means {\it with respect to},
{\it r.h.s} means {\it right-hand side} and {\it l.h.s} means {\it left-hand side}.

\section{Certainty Equivalents and Optimized Certainty Equivalents}
Decision makers are often risk averse when faced with decisions,\footnote{ The St. Petersburg Paradox which is due to  Daniel Bernoulli in 1738 is often mentioned as the first discussion of this topic. For an English translation of the original paper in  Latin see \cite{bern}.}
in particular when monetary rewards or costs have to be optimized.  
Consider for example the following two lotteries:
\begin{itemize}
    \item Lottery 1: receive a reward of $1000$ with probability $0.05$ and $0$ else.
    \item Lottery 2: receive a reward of $50$ with probability 1.
\end{itemize}
Both lotteries have an expected value of $50.$ However when confronted 
with this choice in reality, most people prefer lottery 2, since 
they are risk averse and consider the probability of $0.05$ to be very low. 
Thus, it is reasonable to model risk aversion in decision making. 
This can be done for example by using risk measures.

In what follows let $(\Omega,\mathcal{F},\PP)$ be a probability space. 
All random variables which appear here are defined on this space. 
We will consider Certainty Equivalents and {\em Optimized Certainty Equivalents}. 
Let  $u:\R\to [-\infty,\infty)$ be a strictly increasing, strictly concave utility function. The main purpose
of the utility function is to provide a systematic way to rank alternatives that  captures the principle 
of risk aversion, see, \cite{von2007theory}. This is accomplished whenever the utility function is concave. The
degree of risk aversion  exhibited by the utility function corresponds to the magnitude of the bend
in the function, i.e. the stronger the bend the greater the risk aversion. The degree of risk aversion is 
formally defined by the Arrow-Pratt absolute risk aversion coefficient (\cite{arrow1971theory,pratt1978risk}):
$$\gamma(x):=-\frac{u''(x)}{u'(x)}.$$
Basically, the parameter shows how risk aversion changes with the wealth level. 
Although the actual value of the expected utility of a random outcome is meaningless
except with comparison with other alternatives, there is a derived measure with units that
has intuitive meaning.  
The Certainty Equivalent of a bounded random income $X\in L^\infty(\Omega,\mathcal{F},\PP)$ is defined as
\begin{equation}\label{ce}
 CE(X)= u^{-1} \EE u(X)
\end{equation}
where $\E$ is the expectation operator with respect to the probability measure $\PP.$
$CE(X)$ is the sure amount which yields the same utility as the random outcome. 
The Optimized Certainty Equivalent is defined as follows (\cite{ben2007old}):\\

\begin{definition}
    Let $u:\R\to [-\infty,\infty)$ be a proper, 
closed, concave and non-decreasing utility 
function with $u(0)=0$ and $u'_+(0)\le 1\le u'_-(0)$ 
where $u'_+$ and $u'_-$ are the right and left derivatives 
of $u$.\footnote{Note that $u(x)\ge 0$ for all $x\ge 0$ and 
$u(x)\le x$ for all $x\in\R.$} 
Further let $X\in L^\infty(\Omega,\mathcal{F},\PP)$ be a 
bounded random variable. 
The {\em Optimized Certainty Equivalent}  (OCE) for $X$ is a 
map $S_u:L^\infty(\Omega,{\cal F},\PP) \to \R$ with
$$ S_u(X) = \sup_{\eta\in\R} \{ \eta+\EE u(X-\eta)\}$$
 which is assumed to be a proper function, which means
that the domain $dom S_u:=\{X\in L^\infty(\Omega,{\cal 
F},\PP):\ S_u(X)>-\infty\}$ is not empty and $S_u$ is 
finite on this domain.
\end{definition}
\vspace*{0.4cm}

The interpretation here is that the decision maker may consume 
the amount $\eta$ today and obtain the present value 
$\eta+\EE u(X-\eta)$ as a result. Optimizing over the 
consumption then yields the present value of $X$. 
Among others, $S_u(X)$ has the following properties for $X,Y\in L^\infty(\Omega,\mathcal{F},\PP)$ (see \cite{ben2007old}):
\begin{itemize}
    \item[(P1)] monotonicity: $X\le Y \Rightarrow S_u(X) \le S_u(Y)$;
    \item[(P2)] shift additivity : $ S_u(X+c) = S_u(X)+c$, for any $c\in\R$;
    \item[(P3)] Jensen inequality: $ S_u(X) \le \EE X$;
    \item[(P4)] consistency: $S_u(c)=c$ for any $c\in\R.$
\end{itemize}

Indeed it can be shown that $-S_u$ is a convex risk measure in the sense 
of \cite{follmer2010convex}. A random variable $X$ is now preferred over 
$Y$ if $S_u(X) \ge S_u(Y)$. Thus (P3) and (P4) imply that this preference 
order models {\em risk aversion} since $S_u(X) \le \EE 
X=S_u(\EE X),$ i.e.\ 
the sure amount $\EE X$ is preferred over a random amount $X$ with same expectation. 
Moreover,  it  holds that
$$ \lim_{\delta\to 0} \frac{1}{\delta }S_u(\delta X)= \EE X$$
which  means that the {\em risk-neutral} setting is achieved in the limit.

A further representation of $S_u$ is due to \cite{ben2007old} given by
$$S_u(X)=\inf_{\Q\in\mathcal{Q}}\{ I_\varphi (\Q,\PP) + \EE_{\Q} X\}$$
where $\mathcal{Q}$ is the set of all probability measures $\Q$ absolutely continuous 
w.r.t.\ $\PP$ such that $\frac{d\Q}{d\PP}\in L^1(\Omega,{\cal F}, \PP)$ and $I_\varphi$ is the usual $\varphi$-divergence defined by
$$  I_\varphi (\Q,\PP) = \left\{ \begin{array}{cc}
   \int \varphi\Big(\frac{d\Q}{d\PP}\Big)d\PP,  &  \mbox{ if } \Q\ll \PP\\
   \infty,  &  \mbox{ else.}
\end{array}\right.$$
Here $\varphi:\R \to [0,+\infty]$
is a proper closed convex function with closed interval (containing 1) 
as domain and $\varphi(1)=0$. This representation can be exploited 
in the analysis of risk-sensitive problems and in order to construct 
a connection to robust decision making, see \cite{dai1996connections,bauerle2022distributionally}. 
Indeed this representation consists of the risk-neutral part $ \EE_{\Q} X $ 
where however the infimum over a set $\mathcal{Q}$ of probability measures is taken. 
This resembles   a robust approach. The $\varphi$-divergence term penalizes 
the distance of $\Q$ to $\PP$. The following examples list important special 
cases of the Optimized Certainty Equivalent. \\

\begin{example}\label{ex:differnt:u}
\begin{itemize}
    \item[a)] When we choose $u(t)=\frac1\gamma(1-e^{-\gamma t})$ 
		for $\gamma>0$ we obtain 
   \begin{equation}
       \label{erm}
   S_u(X) = -\frac1\gamma\ln \EE e^{-\gamma X}.
   \end{equation}
The quantity    $-S_u(X)$ is known as {\em entropic risk measure}, see 
    p. 184 in \cite{follmer2010convex}. However, we shall further also refer to (\ref{erm}) as the entropic risk measure.
    It is easy to see that in this case $$ S_u(X)=u^{-1}  \EE u(X)$$ 
		coincides with the Certainty Equivalent of $X$ w.r.t.\ $u$. A
     Taylor series expansion yields
    $$ S_u(X)\approx\EE X- \frac\gamma2 Var(X)$$
which connects the entropic risk to the mean variance criterion.
The entropic risk measure is the most widely used functional which is 
applied in risk-sensitive dynamic decision making. 
This is mainly because it is still mathematically tractable. 
Indeed, the paper of  
\cite{howard1972risk} which is considered to be the first work in this 
field, coined the name {\em risk-sensitive Markov decision process}. 
Since then the adjective 'risk-sensitive' is often used as a synonym for applying the entropic risk measure.
    \item[b)] If for $\alpha\in (0,1)$ we choose
    $$ u(t) = \left\{ \begin{array}{cl}
         \frac{1}{\alpha} t,& t\ge 0 \\
        0, &  t<0
    \end{array}\right.$$
    then $S_u(X)=-CVaR_\alpha(X)$ where  the risk measure {\em Conditional Value-at-Risk} (CVaR) is defined as
$$CVaR_\alpha(X)=\inf_\eta \Big\{\frac{1}{\alpha} \EE (\eta-X)^+ -\eta \Big\}.$$
The Conditional Value-at-Risk is sometimes also called {\em Average Value-at-Risk} or {\em Expected Shortfall.} 
It can be represented as
$$CVaR_\alpha(X) = \frac{1}{\alpha} \int_0^\alpha VaR_\gamma(X)d\gamma $$
where $VaR_\alpha(X)=\inf\{ c\in\R : \PP(X+c<0)\le \alpha\}$ is the Value-at-Risk. In case of a continuous random variable $X$ we also have
$$CVaR_\alpha(X) = \EE[-X | -X \ge VaR_\alpha(X)]. $$
The Conditional Value-at-Risk is not only a convex, but also a coherent risk measure. 
It is the smallest convex risk measure which dominates 
Value-at-Risk, see Remark 4.56 in \cite{follmer2010convex}.
\item[c)] If we choose
$$ u(t) = \left\{ \begin{array}{cl}
t-\frac{1}{2} t^2,& t<1 \\
 \frac{1}{2} , &  t\ge 1
\end{array}\right.$$
then for random variables with $X\le 1+\EE X$ we obtain that $S_u(X)$ is the {\em mean-variance criterion}
$$S_u(X)= \EE X- \frac{1}{2} Var(X).$$
The mean-variance criterion is a popular decision criterion in finance 
since its first appearance in \cite{markowitz1952}. 
However the interpretation is here restricted to random variables with bounded support.
\end{itemize}
\end{example}
\vspace*{0.2cm}

\begin{remark}\label{rem:1}
In what follows we consider optimization problems with rewards. Thus, we maximize $S_u.$ 
In case we want to minimize cost, we have to define the criterion in a different way. In this case let 
$\ell:\R\to (-\infty,\infty]$ be a proper, closed, convex and non-decreasing function bounded from below 
with $\ell(0)=0$ and $\ell'_+(0)\ge 1\ge \ell'_-(0)$. 
For $X\in L^\infty(\Omega,\mathcal{F},\PP)$ 
the {\em Optimized Certainty Equivalent} is then $S_\ell:L^\infty(\Omega,\mathcal{F},\PP) \to \R$ with
$$ S_\ell(X) = \inf_{\eta\in\R} \{ \eta+\EE \ell(X-\eta)\}$$
which is assumed to be a proper function. For $X$ being a cost, this criterion has to be minimized.
\end{remark}

\section{Markov Decision Processes }\label{sec:MDP}
\subsection{The Model}

For dynamic decision making we consider the following controlled Markov process in discrete time, \cite{puterman2014markov,lerma96,bauerle2011markov}. 
\begin{itemize}
    \item[(a)] The state space $E$ is a Borel space (non-empty Borel subset of a Polish space).
    \item[(b)] The action space $A$ is a Borel space.
    \item[(c)] $D\subset E\times A$ is the set of admissible state-action combinations. 
		$D$ contains the graph of a measurable mapping $f:E\to A.$ The sets 
		$D(x)=\{a\in A : (x,a)\in D\}$ of admissible actions in state $x$ are assumed to be compact.
    \item[(d)] $q$ is a regular conditional distribution from $D$ to $E$.
    \item[(e)] The one-stage reward $r:D\to\R_+$ is a
    bounded  Borel measurable function $r(x,a)\le d$ for all $(x,a)\in D$ for some constant $d>0.$ 
\end{itemize}

We define the set of histories of the process. 
At time $k=0$ we have $H_0=E.$ For $k\ge 1$ the set of histories are given by 
$H_k = D^{k}\times E$ and $H_\infty=D\times D\times \ldots $. 
A policy $\pi=(\pi_k)_{k\in\N_0}$ is a sequence of decision rules 
(Borel measurable mappings) from $H_k$ to $A$ such that $\pi_k(h_k)\in D(x_k)$ 
where $h_k=(x_0,a_0,\ldots,x_k)\in H_k.$ The set of all policies is denoted by $\Pi.$ 
Let $F$ be the set of all measurable mappings $f: E\to A$ such that $f(x)\in D(x)$ for every $x\in E.$ By our
assumption $F\not=\emptyset.$ 
A Markovian policy is a sequence $(f_k)_{k\in\N_0}$ where each $f_k\in F.$
The class of Markovian policies is denoted by $\Pi^M$.
A Markovian policy $(f_k)_{k\in\N_0}$ is stationary if there is  some $f\in F$ such that $f_k=f$ 
for every $k\in\N_0,$  i.e.\ 
the same decision rule $f$ is used throughout the time. We identify a stationary policy with 
the element of the sequence. Therefore, the set of all stationary policies will be denoted by $F.$ 
 We have
$$ F \subset \Pi^M \subset \Pi.$$ %

Let 
$(\Omega, \mathcal{F})$ be a measurable space consisting of 
the sample space  $\Omega=(E\times A)^\infty $ with the 
corresponding product  $\sigma$-algebra 
$\mathcal{F}$  on $\Omega.$   
The elements of $\Omega$ are the sequences 
$\omega = (x_0,a_0,x_1,a_1,\ldots) \in H_\infty$ with 
$x_n\in E$ and $a_n\in A$ for $n\in\N_0.$
The random variables $X_0,A_0,X_1,A_1,\ldots$ 
are defined by
$$ X_k(\omega)=x_k,\quad A_k(\omega)= a_k, \qquad k\in\N_0$$
and represent the state and action process, respectively.
Let $\pi\in\Pi$ and the initial 
state $x$ be fixed. Then according to the Ionescu-Tulcea 
theorem there exists a unique probability measure 
$\PP_x^\pi$ on $(\Omega,\mathcal{F})$, which is supported on 
$H_\infty,$ i.e. $\PP_x^\pi(H_\infty)=1.$ Moreover, 
 for $k\in\N:$ 
\begin{itemize}
\item[(a)] $\PP_x^\pi(X_0\in B)=\delta_x(B)$ for all 
$B\in \mathcal{B}(E),$
  \item[(b)] $\PP_x^\pi(A_{k}\in C| h_k) =\delta_{ \pi_k(h_k)}(C)$
for all  $h_k\in H_k,$
 \item[(c)] $\PP_x^\pi(X_{k+1}\in B| x,a_0,\ldots,x_k,a_k) = 
 q(B|x_k,a_k).$
\end{itemize}

\subsection{Risk Neutral Decision Maker}

One of the standard optimization problems for Markov decision processes is 
to find the maximal expected discounted reward:
\begin{align}\label{prob:standard}
    J_\beta^*(x)=\sup_{\pi\in\Pi} J_\beta(x,\pi) 
		\quad \mbox{with}\quad J_\beta(x,\pi)=\EE_x^\pi \left[  \sum_{k=0}^\infty \beta^k r(X_k,A_k) \right]
\end{align}
where $\beta\in[0,1)$ is a discount coefficient and, if possible, an optimal policy $\pi^*$ with $J_\beta^*(x)=J_\beta(x,\pi^*)$. 
Under some continuity and compactness assumptions, the maximal value $J^*_\beta$ and an optimal 
policy can be characterized via the Bellman equation. In order to establish this equation we may use  one of two different sets of conditions which are common in the literature, see
\cite{schal1975dynamic,schal1983stationary}:\\

\noindent
{\bf Condition (S):}
\begin{itemize}
    \item[(a)] The sets $D(x),$ $x\in E,$ are compact.
    \item[(b)] For each $x\in E$ and every Borel set $C\subset E$ the function $q(C|x,\cdot)$ 
		is continuous on $D(x).$
    \item[(c)] The reward $r(x,\cdot)$ is upper semicontinuous on $D(x)$ for each $x\in E.$
\end{itemize}

\noindent
{\bf Condition (W):}
\begin{itemize}
    \item[(a)] The sets $D(x),$ $x\in E,$ are compact and the mapping $x\to D(x)$ is upper semicontinuous.
    \item[(b)] The transition law $q$ is weakly continuous on $D,$ i.e.  the function
    $$ (x,a) \to \int h(y) q(dy|x,a)$$
    is continuous for each continuous bounded function $h.$
   \item[(c)] The reward $r$ is upper semicontinuous on $D$.
\end{itemize}

In what follows let $U(E)$ be the set of all bounded, non-negative upper semicontinuous functions 
on $E$ and $B(E)$ the set of all bounded, non-negative Borel measurable functions on $E.$ We equip these spaces with the supremum norm $\|\cdot\|.$  \\

\begin{theorem}\label{theo:MDP_RN}
    Assume (W) [(S)]. Then
    \begin{itemize}
\item[a)] There exist a unique function $V_\beta\in U(E)$ $[V_\beta\in B(E)]$ and a decision rule 
$f^*\in F$ such that for all $x\in E:$
\begin{align}\label{eq:optimality_rn}
 V_\beta(x)&= \sup_{a\in D(x)}  \Big\{ r(x,a) + \beta \int V_\beta(y) q(dy|x,a) \Big\}\\ \nonumber
 &=r(x,f^*(x)) + \beta \int V_\beta(y) q(dy|x,f^*(x)).
 \end{align}
 \item[b)] Moreover, $V_\beta(x)=J_\beta^*(x)=J_\beta(x,f^*)$ for all $x\in E,$ i.e.\ $f^*\in F$ is an optimal stationary policy.     
    \end{itemize}
\end{theorem}

Theorem \ref{theo:MDP_RN} can be used to establish the link between 
the expected discounted  reward and the long-run average reward defined as:
$$  {\cal J}(x,\pi)=\liminf_{n\to\infty}\frac 1n \EE_x^\pi \left[  \sum_{k=0}^{n-1} r(X_k,A_k) \right]$$
for any initial state $x\in E$ and $\pi\in\Pi.$ The aim is to find a policy $\pi^*\in \Pi$ such that 
${\cal J}(x) := \sup_{\pi\in\Pi}{\cal J}(x,\pi) = {\cal J}(x,\pi^*) $ for every 
$x\in E.$ A first relation between discounted reward and long-run average reward
is provided by the Hardy-Littlewood  theorem. It claims that for bounded sequences 
of real numbers $(R_k)_{k\in\N_0}$ it holds
$$\liminf_{n\to\infty}\frac 1n  \sum_{k=0}^{n-1} R_k\le \liminf_{\beta\to 1} (1-\beta)\sum_{k=0}^{\infty}\beta^k R_k.$$
When we set $R_k:=\E_x^\pi r(X_k,A_k)$ then we immediately obtain
$${\cal J}(x,\pi)\le \liminf_{\beta\to 1} (1-\beta) J_\beta (x,\pi),\quad x\in E,\ \pi\in\Pi,$$
and consequently
$$\sup_{\pi\in\Pi} {\cal J}(x,\pi)\le  \liminf_{\beta\to 1} (1-\beta) J^*_\beta (x),\quad x\in E.$$
A second relation is given via (\ref{eq:optimality_rn}). Let $z\in E$ be a fixed state and put
$h_\beta(x):=V_\beta(x)-V_\beta(z).$ Then simple rearrangements in  (\ref{eq:optimality_rn}) yield 
$$ (1-\beta)V_\beta(z)+h_\beta(x)=\sup_{a\in D(x)}  \Big\{ r(x,a) + \beta \int h_\beta(y) q(dy|x,a) \Big\}, \quad x\in E.$$
Under  certain set of  conditions and 
letting $\beta\to 1,$ the pair $((1-\beta)V_\beta(z), h_\beta(\cdot))$ would converge
to a pair $(\xi,h(\cdot))$ that satisfies the average reward optimality equation
\begin{equation}\label{eq:aoptimality_rn}
\xi+h(x)=\sup_{a\in D(x)}\left\{ r(x,a)+\int h(y)q(dy|x,a)\right\},\quad x\in E.
\end{equation}
If a set of ``reasonably mild'' assumptions is imposed on the family of functions $\{h_\beta(\cdot)\}$ then 
a pair $(\xi,h(\cdot))$ meets the average reward optimality inequality 
\begin{equation}
\label{eq:aioptimality_rn}
\xi+h(x)\le\sup_{a\in D(x)}\left\{ r(x,a)+\int h(y)q(dy|x,a)\right\},\quad x\in E.
\end{equation}
If the maximizer, say $f_*\in F,$ of the r.h.s. in (\ref{eq:aoptimality_rn}) or (\ref{eq:aioptimality_rn}) exists, 
it constitutes an optimal stationary policy, i.e. ${\cal J}(x,f_*)=\sup_{\pi\in\Pi}{\cal J}(x,\pi)$ 
for every $x\in E$ and moreover,
the optimal average reward is independent of the initial state and
$\xi={\cal J}(x,f_*).$ This approach is well-described in the literature. The reader is referred to 
\cite{lerma96,piunovskiy2013} where also other methods are presented with comments and illustrative examples.\\

There are a number of established computational approaches 
which can, often after modifications, also be applied to 
the risk-sensitive cases which we discuss later. 
For example if we consider the setting  of Theorem 
\ref{theo:MDP_RN}, the operator
\begin{align}\label{eq:Toperator_classic}
    Tv(x) := \sup_{a\in D(x)}  \Big\{ r(x,a) + \beta \int v(y) q(dy|x,a) \Big\}
\end{align} 
is a contraction on a suitable function space into the same 
function space. Then applying 
Banach's fixed point theorem, the value function and the 
optimal policy  can be approximated by iterating the 
$T$-operator. Alternatively, one can start with an arbitrary 
stationary policy, given by a decision rule $f\in F$, compute 
the corresponding value $V_f(x)=J_\beta(x,f)$ (see 
\eqref{prob:standard}) and improve it by computing the maximum 
points on the r.h.s.\ of \eqref{eq:Toperator_classic} 
with $v$ replaced by $V_f.$ Under mild assumptions this 
procedure converges to the optimal solution. For computational 
purposes it is often more convenient to consider the so-called 
Q-function, which is defined as follows
\begin{align*}
    Q(x,a) := r(x,a) + \beta \int V_\beta(y) q(dy|x,a).
\end{align*} 
Note that we have $V_\beta(x) =  \sup_{a\in D(x)}  Q(x,a)$ and 
\begin{align*}
    Q(x,a) = r(x,a) + \beta \int \sup_{a'\in D(y)} Q(y,a') q(dy|x,a).
\end{align*} 
This representation has the advantage that the maximization can 
be done before the integration. The algorithms discussed so far 
are only applicable when the state and action spaces are of low 
dimension and all data of the model  are known. Modern 
approximate solution techniques are summarized under the name 
{\it  Reinforcement Learning} (RL). 
The aim of these methods is to 
find an optimal strategy while simultaneously learn the right 
model. 
A popular approach is Q-learning, where the learned action-value function $Q^{(t)}$ directly approximates the Q-function. First we initialize  $Q^{(0)}$ 
arbitrarily. Then we repeat the following steps:
\begin{enumerate}
    \item Choose an admissible pair $(x,a)$ at random and observe the next state $y$ (or generate $y\sim q(\cdot|x,a)$).
    \item Update at $(x,a):$
    \begin{align*}
        Q^{(t+1)}(x,a) := (1-\alpha_t) Q^{(t)}(x,a)+ \alpha_t \Big(r(x,a)+\beta \sup_{a'\in D(y)} Q^{(t)}(y,a'). \Big)
    \end{align*}
    where the learning rates $(\alpha_t)$ have to be chosen appropriately.
\end{enumerate}
Under mild assumptions this method is known to converge to the 
$Q$-function. Further methods parametrize the class of policies 
and thus, the value function and estimate the optimal 
parameters. Though not being optimal, in this situation it is 
more convenient to work with randomized policies.  In order to 
find the best parameters in this setting, often the gradient is 
computed and parameters are updated by a gradient ascent rule. 
For computational issue consult  among others with
\cite{sutton2018reinforcement,powell2022reinforcement,hambly2023recent}.

As discussed in the previous section this criterion does not account for deviations around the mean or in other words the risk of the decision maker. Thus, in what follows we consider risk-sensitive optimization criteria.

\section{Markov Decision Processes with Recursive Risk-Sensitive Preferences} \label{s:mdp_rec}

Measuring risk in a stochastic dynamic process is much more complicated 
than in a single-step situation. It may be measured at every stage and 
then aggregated or measured by a nested application of risk measures 
or a single-step risk measure is applied to the aggregated discounted reward.

In what follows we concentrate on the underlying controlled 
stochastic dynamic process 
to be Markovian (like in the previous section) 
and that the Optimized Certainty Equivalent risk measures 
are applied recursively. This setting guarantees that the 
optimality principle holds 
and optimal policies are stationary.

For $k\in\N_0$ let 
$$B(H_k):=\{ v:H_k\to\R_+ : v \mbox{ is measurable, 
bounded}\}$$ 
be equipped with the supremum norm $\|\cdot\|.$ Let $\pi=(\pi_k)_{k\in\N_0}\in\Pi$ be an arbitrary policy. 
For $v_{k+1}\in B(H_{k+1})$ and $h_k\in H_k$ we define a conditional Optimized Certainty Equivalent
\begin{eqnarray*}
    \lefteqn{S^{(x_k,\pi_k(h_k))}_u\big(v_{k+1}(h_k,\pi_k(h_k),X_{k+1})\big) :=}\\& 
		&\sup_{\eta\in\R} \Big\{ \eta+\int u(v_{k+1}(h_k,\pi_k(h_k),y)-\eta) q(dy|x_k,\pi_k(h_k))\Big\}
\end{eqnarray*}
 where the random variable $X_{k+1}$ has the distribution $q(\cdot|x_k,\pi_k(h_k)).$ 
Then we define the operator $L_{\pi_k}$ as follows:
\begin{eqnarray*}
    (L_{\pi_k} v_{k+1})(h_k) &=& L_{\pi_k} v_{k+1}(h_k) := \\ && r(x_k,\pi_k(h_k))+\beta  
		S^{(x_k,\pi_k(h_k))}_u\big(v_{k+1}(h_k,\pi_k(h_k),X_{k+1})\big) \nonumber
\end{eqnarray*} 
where $\beta\in[0,1)$ is a discount factor. The operator $L_{\pi_k}$ is monotone by (P1), i.e. 
$$ v_{k+1}\le w_{k+1} \Rightarrow L_{\pi_k} v_{k+1} \le L_{\pi_k} w_{k+1}$$
for $v_{k+1},w_{k+1} \in B(H_{k+1})$. By (P1) and (P4) it holds 
\begin{align}\label{eq:bound_L}
    0\le L_{\pi_k} v_{k+1}(h_k) \le d+\beta \|v_{k+1}\|.
\quad \mbox{for any}\quad h_k\in H_k.\end{align}

Let now $N\in\N$. For the $N$-stage decision model we apply these operators recursively. Thus, for 
an initial state $x\in E$, the total discounted recursive risk-sensitive reward under policy $\pi$ is given by
$$ J_N(x,\pi)= (L_{\pi_0}\circ \ldots \circ L_{\pi_{N-1}})\mathbf{0}(x) $$
where $\mathbf{0}$ is the function $\mathbf{0}(h_k)\equiv 0$ for all $h_k\in H_k, k\in\N_0.$ 
For $N=2$ this equation reads
\begin{align*}
    J_2(x,\pi) &= (L_{\pi_0}\circ L_{\pi_1})\mathbf{0}(x) = L_{\pi_0} (L_{\pi_1}\mathbf{0} )(x)\\
    &=  r(x,\pi_0(x))+\beta  S^{(x,\pi_0(x))}_u \big(r(X_1,\pi_1(x,\pi_0(x),X_1))\big).
\end{align*}
Aggregation over time is still additive in this approach. 
By our assumptions and (P1), the sequence $(J_N(x,\pi))_{N\in\N}$ is 
non-decreasing and bounded from below by $0$ for all $x\in E$ and $\pi\in\Pi.$ 
Moreover, by \eqref{eq:bound_L} we obtain
$$ J_N(x,\pi) \le \frac{d}{1-\beta},\quad x\in E,\ \pi\in\Pi,\ N\in\N.$$
Hence the limit $\lim_{N\to\infty} J_N(x,\pi)$ exists for $x\in E$ and $\pi\in\Pi.$\\

\begin{problem}
    For an initial wealth $x\in E$ and a policy $\pi\in\Pi$ 
		we define the total discounted recursive risk-sensitive reward by
    $$ J(x,\pi) := \lim_{N\to\infty} J_N(x,\pi).$$
    The aim of the decision maker is to find the maximal value, i.e.
    $$ J^*(x) := \sup_{\pi\in\Pi} J(x,\pi),\quad x\in E$$
    and a policy $\pi^*$ such that  $J(x,\pi^*)=J^*(x),$ $x\in E.$
\end{problem}
\vspace*{0.2cm}

In order to solve the problem we use dynamic programming. 
We need essentially the same assumptions as in the risk-neutral case.

A proof of the following theorem can be found in the appendix.\\

\begin{theorem}\label{theo:prob1}
Assume (W) [(S)]. Then
\begin{itemize}
        \item[a)] There exist a unique function $V\in U(E)$ $[V\in B(E)]$ 
				and a decision rule $f^*\in F$ such that for all $x\in E:$
        \begin{align}\label{eq:optimality}
            V(x)&= \sup_{a\in D(x)} \Big\{ r(x,a) + \beta S^{(x,a)}_u(V(X_1)) \Big\}\\ \nonumber
            &= r(x,f^*(x)) + \beta S_u^{(x,f^*(x))}(V(X_1))
        \end{align}
        where $S_u^{(x,a)}$ indicates that $X_1$ has the distribution $q(\cdot|x,a).$
         \item[b)] Moreover, $V(x)=J^*(x)=J(x,f^*)$ for all $x\in E,$ 
				i.e.\ $f^*\in F$ is an optimal stationary policy.     
    \end{itemize}
\end{theorem}

%

If $u$ is an exponential utility then we obtain in the previous case that 
$S_u$ is the entropic risk measure (Example \ref{ex:differnt:u} a)) and the  
optimality equation \eqref{eq:optimality} reduces to (see \cite{asienkiewicz2017note})
\begin{align}\label{eq:entropic_rec}
  V(x)= \sup_{a\in D(x)} \Big\{ r(x,a) -  \frac\beta\gamma \ln \Big\{ \int \exp(-\gamma V(y)) q(dy|x,a)\Big\} \Big\}  
\end{align}
for $x\in E.$
The expression in brackets on the r.h.s is also referred to as {\em risk-sensitive Koopmans operator} (see 
\cite{miao2020economic,stachurski23}). By applying the exponential function on both sides, 
the equation can also be written as
\begin{align}\label{eq:entropic_rec2} 
\tilde V(x)= \sup_{a\in D(x)} \Big\{ e^{-\gamma r(x,a)} 
\left(  \int \tilde V(y)q(dy|x,a)\right)^{\beta} \Big\} 
\end{align}
with $\tilde V(x)= e^{-\gamma V(x)}$ which yields a multiplicative Bellman equation.

A discounted, recursive entropic cost linear quadratic Gaussian regulator 
problem with the infinite time horizon has been treated in \cite{hansen1995discounted}. 
Conditional consistency of the recursive entropic risk measure is discussed in \cite{dowson2020multistage}. 
An efficient learning algorithm for recursive Optimized Certainty Equivalents 
based on value iteration and upper confidence bound 
can be found in \cite{xu2023regret,fei2021exponential} 
where the latter concentrates on the entropic risk measure.

CVaR optimization (which is according to Example \ref{ex:differnt:u} b) another special case) 
for a finite time horizon applied at the terminal wealth has been considered in 
\cite{rudloff2014time,pflug2016time} and for the infinite time horizon in \cite{uugurlu2018robust}. 
The authors also discuss time-consistency issues of optimal policies.
\cite{shapiro2013risk} consider risk averse approaches (in terms of a weighted 
criterion of expectation and CVaR) to multistage (linear) stochastic programming problems based 
on the Stochastic Dual Dynamic Programming method. For further computational approaches 
see \cite{kozmik2015evaluating}.
The recursive CVaR is very popular for applications (see Section \ref{sec:appl}).  

Some papers have studied the more general class of convex risk measure for a 
nested application to stochastic dynamic decision problems. For example 
\cite{shen2013risk,shen2014unified,chu2014markov,bauerle2022markov} consider 
the infinite time horizon, unbounded cost functions and establish optimality equations 
and existence of optimal policies. \cite{martyr2022discrete}
consider
an iterated $\mathbb{G}$-expectation for non-Markovian optimal 
switching problems.
In \cite{dowson2022incorporating} the problem is tackled as a multistage stochastic program. 
Algorithms based on stochastic dual dynamic programming  and the special role of the entropic 
risk measure in this class are discussed in \cite{shapiro2021tutorial,dupavcova2015structure}. 
\cite{philpott2013solving} 
use inner, outer approximations based on dynamic programming.  
Further algorithms can be found in \cite{le2007robust,tamar2016sequential,
guigues2016convergence,huang2021convergence}. Algorithms for a finite time horizon 
and convex risk measures based on reinforcement learning are 
studied in \cite{coache2023reinforcement}.
 
There are further recursive risk-sensitive preferences in the literature which are not covered by our model. 
\cite{kreps1978} and \cite{epstein2013substitution} 
propose an alternative specification of lifetime value that separates and independently parametrizes
temporal elasticity of substitution and risk aversion. To be more precise 
\cite{kreps1978} consider a finite time horizon recursive preferences with the
conditional Certainty Equivalent\footnote{We mean here (like in the case of a conditional OCE) a  Certainty Equivalent  
that maps a random variable that is measurable with the next 
period's information into a random variable that is 
measurable with respect to the current period's information. 
} defined with $u(x)=x^{1-\gamma},$ $\gamma>0$ and $\gamma\not=1$, see \eqref{ce}. Here, the parameter $\gamma$ is responsible for the level of relative risk aversion.
\cite{epstein2013substitution} generalize their approach to 
the infinite time horizon and 
suggest the following  form of aggregation:
$$ v_n(x) := \left( (1-\beta) (r(x,f_{n}(x)))^{1-\rho}  + 
\beta \left(\int v_{n+1}(y)^{1-\gamma} q(dy|x,f_{n}(x))\right)^{\frac{1-\rho}{1-\gamma}}\right)^{\frac 1{1-\rho}}. $$
The function  $v_n$ denotes the future payoff from period $n\in \N_0$ onwards when 
the process is governed by a Markovian policy $(f_n)\in \Pi^M.$ Moreover, we assume that $\rho >0$
and $\rho\not=1.$
The value $1/\rho$ represents  a {\em Constant Elasticity of Intertemporal Substitution} (CES). Therefore, the Epstein-Zin aggregator (named from their authors)
is also called a CES time aggregator. \cite{epstein2013substitution} obtain a remarkable result for the existence of recursive utilities across the broad set of parameters $\gamma$ and $\rho.$ Their results 
have been further strengthened by \cite{ozaki1996dynamic}
who provide an extensive analysis of existence and uniqueness of recursive utilities by introducing the notion of biconvergence. This concept requires that returns can be sufficiently discounted from above and sufficiently discounted from below. Moreover, their results are useful for studying dynamic programming with non-additive stochastic objectives in a pretty general setting. 
The Epstein-Zin time aggregator has  also been examined by \cite{weil1993precautionary} but with
the conditional Certainty Equivalent defined by  an exponential 
utility function. The function $v_n$ is there given as 
follows
\begin{eqnarray*}    
\lefteqn{v_n(x) :=}\\&&
\left( (1-\beta) (r(x,f_{n}(x)))^{1-\rho}  + 
\beta \left(-\frac 1\gamma\ln\int \exp(-\gamma v_{n+1}(y)) q(dy|x,f_{n}(x))\right)^{1-\rho}\right)^{\frac 1{1-\rho}}. 
\end{eqnarray*}
The aforementioned recursive preferences are very popular among economists 
(see for instance \cite{stachurski23,miao2020economic} and references cited therein) who put a lot of criticism 
on the standard expected discounted utility. To learn more on this subject the reader 
is referred to the notes following Chapter 7 in  \cite{stachurski23}.  It is worthy to mention that the CES 
time aggregator
and different conditional Certainty Equivalents have been also exploited  within dynamic programming framework by a number of authors, see the references in
\cite{ren2018dynamic}, Chapter 8 in \cite{stachurski23}.


 \cite{marinacci2010unique} propose a new class of Thompson aggregators and study a class of quasi-arithmetic Certainty Equivalent operators that generalize those of \cite{kreps1978}. Based on specific properties of such operators and the time aggregator they provide a comprehensive analysis of 
existence, uniqueness  and global 
attractivity of  a continuation value process. 
Particularly, they make use of monotonicity and concavity of the Thompson aggregator and subhomogeneity of the quasi-arithmetic operator. These facts allow them to define a contraction within the Thompson metric.  

 \cite{bloise2018convex} develop an  approach to convex 
programs for bounded recursive utilities. Their technique relies
upon the theory of monotone concave operators. An extension is given in \cite{bloise2021not}. 
\cite{iwamoto1999conditional}, on the other hand, 
treats optimization problems with nested recursive utilities given 
by applying appropriate functions. A dynamic programming approach is 
used to solve the problems.

Further extensions include \cite{feinstein2017recursive} and \cite{schlosser2020risk}. 
In the latter paper 
a multi-valued dynamic programming approach is considered that allows 
to control the moments of the distributions of future rewards. 
The former paper is devoted to the development of 
set-valued risk measures and the recursive algorithms for a dynamic setting.

\section{Markov Decision Processes with Risk-Sensitive Discounted Reward }\label{sec:outside}

Instead of applying the Optimized Certainty Equivalent 
recursively  one can also apply it to the discounted sum of 
the rewards. Within such a framework the optimal policies 
need  not be time-consistent. 
We say that a multiperiod stochastic decision problem is time-consistent, if resolving the problem at
later stages (i.e., after observing some random outcomes), the original solutions remain
optimal for the later stages.
For a recent survey of different approaches to  dynamic 
decision problems with risk measures  and their connection 
to time-consistency, see \cite{homem2016risk}. 
We only mention here a stream of references \cite{kreps1977decisionI,kreps1977decision,
iwamoto2004stochastic,pflug2005measuring,
pflug2006value,
ruszczynski2010risk,osogami2011iterated,shapiro2012minimax,
philpott2013solving} that contributed to this issue among others. Below we provide a simple example that illustrates the problem of time-consistency in the approach taken in this section.

We use the same MDP model as in the previous 
section. For fixed history 
$\omega=(x_0,a_0,x_1,a_1,\ldots)\in H_\infty$ 
let us define the sum of the discounted rewards by
$$
    R_\beta^\infty(\omega) := \sum_{k=0}^\infty \beta^k r(x_k,a_k)
$$
where we always assume that the initial state $x_0=x.$ We also put
\begin{equation}\label{eq:oce}
 S_u^\pi(R_\beta^\infty)=\sup_{\eta\in\R}
		\Big\{ \eta+\int_{H_\infty} u(R_\beta^\infty(\omega)-\eta) \PP^\pi_x(d\omega)\Big\},
\end{equation}
		where with a little abuse of notation $R_\beta^\infty$ in \eqref{eq:oce} 
		is now understood as a random variable on $(\Omega,{\cal F})$ with the distribution $\PP_x^\pi$
  supported on $H_\infty.$
In other words,  $S_u^\pi$ indicates that the distribution of $ R_\beta^\infty$ is  $\PP_x^\pi.$
Then we consider the following problem.\\

\begin{problem}\label{prob:outside}
    For initial wealth $x\in E$ and policy $\pi\in\Pi$ we define  
		the total discounted  risk-sensitive reward by
    $$ J(x,\pi) := S_u^\pi(R_\beta^\infty)$$
    The aim of the decision maker is to find the maximal value, i.e.
    $$ J_\infty (x) = \sup_{\pi\in\Pi} S_u^\pi(R_\beta^\infty),\quad x\in E$$
    and a policy $\pi^*\in \Pi$ such that  $J(x,\pi^*)=J_\infty(x),$ $x\in E.$
\end{problem}
\vspace*{0.2cm}

A comparison between the obtained values  
when a coherent risk measure is applied outside or recursively (without control problem), 
can be found in \cite{iancu2015tight}.  
Note that in case of no discounting ($\beta=1$) Problem 1 and Problem 2 
are equivalent. This follows from (P2) and (P4). 
However, discounting ensures that the value of the problem is finite since we have bounded rewards. 
Without discounting it depends on the distribution of $(X_k,A_k)_k$ whether the expectations are finite. 
The motivation or interpretation of applying the risk measure outside is 
somewhat easier than for the recursive application of the risk measure. 
It can be deduced in particular from the different representations of $S_u$ in Example \ref{ex:differnt:u}.

In order to solve Problem \ref{prob:outside} note that by definition of $S_u^\pi$
\begin{align}\label{eq:innerouter}
 \sup_{\pi\in\Pi} S_u^\pi(R_\beta^\infty) &= 
\sup_{\pi\in\Pi}  
		\sup_{\eta\in\R} \Big\{ \eta+ \E_x^\pi [u(R_\beta^\infty -\eta)]\Big\}\\& \nonumber
    =   \sup_{\eta\in\R} \Big\{ \eta+  \sup_{\pi\in\Pi}  \E_x^\pi [u(R_\beta^\infty -\eta)]\Big\}.
\end{align}
Thus, we essentially have to solve $\sup_{\pi\in\Pi} 
 \E^\pi [u(R_\beta^\infty-\eta)]$ first. The challenge here is that there is no obvious 
optimality equation for solving the problem. A way to work around 
this is to enlarge the state space.  
This has been done in \cite{bauerle2014more}. 
More precisely, it is helpful to introduce a new MDP on an extended state space
$\widetilde E := E\times [-\eta,\infty) \times [0,1].$ 
Decision rules $f$ are now measurable mappings from 
$\widetilde E$ to $A$
 respecting $f(x, y, z)\in D(x)$ for every $(x,y,z)\in\widetilde E.$ 
 Denote this set of decision rules by $\widetilde F.$ 
Policies are defined in an obvious way and with a little abuse of notation denote the set of all policies in this new MDP by $\Pi$.  
 For any policy $\pi\in\Pi$
 let
\begin{align}\nonumber
    V^\pi_{\infty}(x,y,z) &:= \E^\pi_x [u(z R_\beta^\infty +y)],\\ \label{eq:probh1}
     V_{\infty}(x,y,z) &:= \sup_{\pi\in\Pi}  V^\pi_{\infty}(x,y,z)
\end{align}
be the value functions on an extended state space.
Thus, we are looking for $V_\infty(x,-\eta,1)$ which is the value of 
the inner optimization problem in \eqref{eq:innerouter}. Let us denote
$U(\widetilde E)$ to be the set of all upper semicontinuous functions $v$ 
with $v(x,\cdot,\cdot)$ is continuous and increasing in both variables 
for all $x,$  and $v(x,y,z)\ge u(y).$
Moreover, denote
$$ \overline b(y,z) := u(z d/(1-\beta) +y),\quad \underline b(y,z) := u(z \underline{d}/(1-\beta) +y)$$
where $\underline{d}$ is a lower bound for $r$ (possibly zero).
The next theorem summarizes the solution.\\

\begin{theorem}\label{theo:prob2}
    Assume (W). Then
    \begin{itemize}
        \item[a)] There exist a unique function $V\in U(\widetilde E)$ with 
				$\underline b \le V\le \overline b$ and a decision rule $\widetilde f^*\in \widetilde F$ 
				such that for all $(x,y,z)\in \tilde E:$
        \begin{align*}
            V(x,y,z)&= \sup_{a\in D(x)} \Big\{  \int V(x',z r(x,a)+y,z\beta) q(dx'|x,a) \Big\}\\
            &=  \int V(x',z r(x,\widetilde f^*(x,y,z))+y,z\beta) q(dx'|x,\widetilde f^*(x,y,z)).
        \end{align*}
 Moreover, $V(x,y,z)=V_\infty(x,y,z)$ for every $(x,y,z)\in\widetilde E.$ 
         \item[b)] There exist an optimal $\eta^*$ in \eqref{eq:innerouter} 
				and a policy $\pi^*=(g^*_0,g^*_1,\ldots)$ with
         $$ g_n^*(h_n) = f^*\Big(x_n, \sum_{k=0}^{n-1} \beta^k r(x_k,a_k)-\eta^*,\beta^n\Big).$$
Moreover, $\pi^*$ is an optimal  policy for Problem \ref{prob:outside}.     
    \end{itemize}
\end{theorem}

If we denote the operator $T: U(\tilde E) \to U(\tilde E)$ by 
$$Tv (x,y,z) :=  \sup_{a\in D(x)} \Big\{  \int v(x',z r(x,a)+y,z\beta) q(dx'|x,a) \Big\},$$ 
then it can also be shown that $ T^n \underline{b} \uparrow V_\infty $ and 
$T^n \overline{b} \downarrow V_\infty$ for $n\to\infty.$ This implies 
that value iteration works here and yields numerical bounds on the value function. 
In \cite{bauerle2014more} it has also been shown that the policy improvement converges.

If $u$ is an exponential utility we obtain in the previous case that $S_u^\pi$ 
is related to the entropic risk measure (Example \ref{ex:differnt:u} a)). Here we can drop the component $y$ and obtain $J_\infty(x)=V_\infty(x,1)$ where $V_\infty(x,z)=\sup_{\pi\in\Pi} 
S_u^\pi(z R_\beta^\infty)$ satisfies in this case
\begin{align*}
    V_\infty(x,z)= \sup_{a\in D(x)} \Big\{  z r(x,a) - 
		\frac1\gamma \ln \int \exp(-\gamma V_\infty(x',z\beta)) q(dx'|x,a) \Big\}.
\end{align*}  
Note here the difference to the optimality equation given in \eqref{eq:entropic_rec} 
where we use the nested application of the entropic risk measure. In case $\beta=1$ 
the value function 
$V_\infty$ does not depend on $z$ and both equations coincide. 

Next we give a simple example from \cite{jaquette1976utility} to show the difference
in optimal policies within 
 the aforementioned frameworks.\\

\begin{example} Let us consider an MDP model with 
$E=\{1,2,3\},$ $A=\{a,b_1,b_2\}.$ The decision maker has 
only a choice in state $x=1,$ namely $D(1)=\{b_1,b_2\}.$
In addition, $D(2)=D(3)=\{a\}.$
The transition probabilities are:
$$q(2|1,b_1)=1-q(3|1,b_1)=0.5,\quad q(2|1,b_2)=1-q(3|1,b_2)=0.9.$$
From state 2 and from state 3 the process always
jumps to state 1 with probability 1. The rewards are 
as follows:
$$r(1,b_1)=0,  \quad r(1,b_2)=1, \quad r(2,a)=0, \quad r(3,a)=8.$$
Obviously, there are two stationary strategies $f$ and $g,$ i.e. $f(1)=b_1,$ $g(1)=b_2$ and $f(2)=g(2)=f(3)=g(3)=a.$
Assume that $\beta=1/2$ and the initial state is $x_0\equiv 1.$ 
Then, the decision maker essentially chooses
between two independent gambles every other period.
The first gamble, call it $X_{f},$ gives the payoff $0$ or $4$ with 
equal probabilities whilst the second gamble, call it $X_{g},$ yields
the reward $1$ with probability $0.9$ or $5$ with 
probability $0.1.$ Since 
$\E  X_{f}=2>\E  X_{g}=1.4,$ the risk-neutral decision maker prefers a stationary policy 
$f$ to $g$. Hence, the maximal expected discounted reward is equal to
$$J_{1/2}(1)= \sum_{n=0}^\infty\left(\frac 1 2\right)^{2n}\E X_{f}= 8/3\approx 2.6666.$$

Let us suppose that the decision maker uses the Optimized Certainty Equivalent defined in \eqref{erm}
with $\gamma=1.$ Consider first  Problem 1. Then, Theorem \ref{theo:prob1} takes the following form
$$V(1)=\max\left\{-\frac 12\ln\left(\frac 12e^{-V(2)}+\frac 12e^{-V(3)}\right),
1-\frac 12\ln\left(\frac 9{10}e^{-V(2)}+\frac 1{10}e^{-V(3)}\right)\right\}$$
and
$$V(2)=\frac{V(1)}2,\quad V(3)=8+\frac{V(1)}2.$$
Then, $g$ is an optimal stationary policy and the maximal reward is 
$$V(1)=\frac{4}{3}\left(1+\ln\left(\sqrt{\frac{10}{9+e^{-8}}}\right)\right)\approx 1.4035.$$

Now let us turn to Problem 2. In our case the aim is to maximize over the set of all policies $\pi\in\Pi$ 
the functional
$$J(1,\pi) =-\ln \E_1^\pi e^{-\sum_{k=0}^\infty (1/2)^k r(X_k,A_k)}.$$
This is equivalent to minimization of the expression
$\bar{J}(1,\pi)=\E_1^\pi e^{-\sum_{k=0}^\infty 
(1/2)^k r(X_k,A_k)}$ 
over the set of all $\pi\in\Pi.$ Since the decision 
maker chooses in each period between two independent 
gambles $X_f$ and $X_g,$ then 
$$\bar{J}(1,\pi)=\E^\pi_1 \exp\left\{-\sum_{n=0}^{\infty}\left(\frac 12\right)^{2n} X_{2n} \right\},$$
where $X_0,X_2,\ldots$ are independent random 
variables with the distribution as $X_f$ or $X_g,$
depending whether the policy $\pi=(\pi_k)$  indicates to use  $f$ or $g$ in  period $k=0,2,4,\ldots$. Clearly, $\pi_k\equiv a$ for $k=1,3,5,\ldots.$
Therefore,
$$\bar{J}(1,\pi)=\prod_{n=0}^\infty \E
e^{-(1/2)^{2n} X_{2n}}.
$$
Observe that
$$\E e^{-s X_{f}} > \E e^{-s X_{g}} \quad \iff\quad \frac 12+\frac 12 e^{-4s}>\frac 9{10}e^{-s}+\frac 1{10} e^{-5s}.$$
This holds for $s>0.455904.$ Hence, for the decision 
maker $g$ is better than $f$ in periods $2n$ for which  $(1/2)^{2n}>0.455904.$ This is
equivalent to $2n<1.1332.$ 
Summing up, the optimal policy is $(g,f,f,f\ldots).$
 Obviously, the policy is not stationary and it is
 not time-consistent \footnote{To be more precise, we have no time-consistency within the class of policies where decisions are only based on the current wealth. However, when we consult Theorem \ref{theo:prob2}, we see that there is some stationarity of the optimal policy on the extended state space.}. However, this policy is 
 ultimately stationary, i.e., there is a period  
 such that from this period onwards the policy is 
 stationary. In fact, \cite{jaquette1976utility} proves that an MDP with a finite state space and the entropic risk measure must be ultimately stationary.
 This has not to be true for MDPs with an infinite state space.
 For other examples illustrating the lack of stationarity and time-consistency the reader  is referred to \cite{brau1998controlled}. 
\end{example}
\vskip 4mm

The first studies of this entropic setting are due to \cite{howard1972risk} 
and \cite{jaquette1976utility}. 
Linear-quadratic problems with a finite time horizon  and the entropic risk 
measure are considered 
in \cite{jacobson1973optimal,whittle1981risk}. A more general approach can be 
found in 
\cite{chung1987discounted} where fixed point theorems for the whole distribution of 
 the infinite time horizon discounted reward in a finite MDP are considered. 
In \cite{collins1998finite} the authors deal with a finite time horizon problem where 
they maximize  a strictly concave functional of the distribution of the terminal state.
 \cite{coraluppi1999risk} connect the problem with the  entropic risk measure 
to a minimax criterion for finite state MDPs. A turnpike theorem 
for a risk sensitive  MDP model with stopping is shown in \cite{denardo2006turnpike}. Though 
 \cite{di1999risk} consider the average reward criterion, they also 
solve as a by-product the infinite time horizon discounted model with Borel state and action spaces.

Numerical methods for the  MDP with the entropic risk measure and finite and infinite time 
horizons are given in \cite{hau2023entropic}. 
A finite time horizon non-discounted MDP with Borel state and action spaces 
and with entropic risk measure is considered in \cite{chapman2021classical}. 
General Certainty Equivalents for MDPs with Borel state and action spaces and 
finite and infinite time horizons are treated in \cite{bauerle2014more}.  
Partially observable MDPs with the entropic risk measures are examined in 
\cite{james1994risk,fernandez1997risk,bauerle2015,bauerle2017partially}.

The special case of optimizing the CVaR of $R^\infty_\beta$ 
with bounded reward has been considered in \cite{bauerle2011avar}. 
A numerical algorithm and the connection to robust optimization problems is 
discussed in \cite{chow2015risk,ding2022sequential}.  Unbounded cost problems 
with CVaR are treated in \cite{uugurlu2017controlled}.
In \cite{chapman2022optimizing} the authors minimize the CVaR of a maximum 
random cost over a finite time horizon. 
\cite{kadota2006discounted}  maximize the expected 
utility of the total 
discounted reward  subject to multiple expected 
utility constraints.

\section{Markov Decision Processes with Other Risk-Sensitive Payoff Criteria }\label{sec:other}

In this section we focus on other payoff criteria 
than those considered in Sections 
\ref{s:mdp_rec} and \ref{sec:outside}. 
We start with average risk-sensitive payoff criteria
when a controller is equipped with a constant 
Arrow-Pratt's risk coefficient, 
i.e. she evaluates her future income using an exponential utility function. 
However, sometimes instead of a reward $r$ in the MDP we shall study a cost $c: D\to\R_+.$
This is because the papers published so far with this criterion mainly deal with a minimization problem and
moreover,  the cost minimization is not equivalent to the reward maximization 
when changing the sign in the cost function
as in the risk-neutral case (see also Remark \ref{rem:1}).\\

\begin{problem}
For an initial state $x\in E$ and a policy $\pi\in\Pi$
we shall consider the following cost functional:
$${\cal J}(x,\pi)=\limsup_{n\to\infty} \frac 1{\gamma n}\ln \EE_x^\pi\left[\exp\left(\sum_{k=0}^{n-1}
\gamma c(X_k,A_k)\right)\right]$$     
for $\gamma>0.$ 
\end{problem}

Here in order to ensure that the average risk-sensitive cost is well-defined,
let us assume as before  that $c$ is bounded.  The objective is to find the minimal cost
$$\xi(x):=\inf_{\pi\in\Pi} {\cal J}(x,\pi).$$
 The policy $\pi^*$ is optimal for the {\em ergodic risk-sensitive control problem} if
$${\cal J}(x,\pi^*)=\inf_{x\in E} \xi(x), \quad x\in E.$$
Note that then the optimal cost $\xi(x)$ must be independent of $x.$

The paper of 
\cite{howard1972risk}\footnote{In their paper the maximization problem is studied.} 
is a pioneering work that deals with the 
aforementioned problem for MDPs with finite state and action spaces. They assume that the Markov chain is aperiodic 
and comprises one communicating class under any stationary policy. A Perrron-Frobenius theory of positive matrices
allows them to establish a solution to the optimality equation which is of the form
\begin{equation}\label{eq:ergodic}
\xi_o+h(x)=\min_{a\in D(x)}\left\{ c(x,a)+\frac 1\gamma\int \exp(\gamma h(y))q(dy|x,a)\right\}
\end{equation}  
for every $x\in E.$ Here $\xi_o$ is a real number and $h: E\to\R$ is a given function. 
If the equation holds, it is possible to prove two points. Firstly, 
the optimal cost is $\xi(x)=\xi_o/\gamma$ for every $x\in E.$
Secondly, the minimizer of the r.h.s. in \eqref{eq:ergodic} (if exists), say $f_*,$ defines an optimal 
stationary policy $f_*\in F,$ which means $ \frac{\xi_o}\gamma={\cal J}(x,f_*),$ $x\in E.$
It should be noted that the optimal cost need not be constant (unlike in the risk neutral case) 
if the Markov chain induced by a stationary policy has transient states, consult with \cite{brau1998controlled} for
counterexamples. The communication properties of the Markov chains in the analysis of the ergodic 
risk-sensitive control problem are underlined in \cite{cavazos2002alternative}.  
Since then the finite state space models have been extensively developed and
the Perron-Frobenius theory has been employed, see among others \cite{sladky2018,sladky2008growth,
rothblum1984multiplicative,cavazos2009perron} 
and references cited therein.
In addition, the Perron–Frobenius theory provides a link
between risk-sensitive control and the Donsker–Varadhan theory of large
deviations. It is known that, under suitable recurrence conditions,
the occupation measure of a Markov process satisfies the large deviation
principle with rate function given by the convex conjugate of a long run
expected rate of exponential growth function. 
Such a variational formula  for  the optimal growth rate of reward in the spirit of the Donsker–Varadhan formula
is given in \cite{anantharam2017variational} where the 
existence of a Perron–Frobenius eigenvalue and an associated eigenfunction is analyzed by the nonlinear Krein–Rutman theorem. For further results in this direction the reader is referred to 
\cite{cavazos2018characterization,arapostathis2016}.

A nice characterization of an optimal cost via a minimization problem in a finite dimensional Euclidean space 
is given in \cite{cavazos2005characterization} where the transition law of the Markov chain satisfies a simultaneous Doeblin condition. This result is generalized to an MDP model on a Borel state space in \cite{cavazos2010discounted}.

The second approach for solving ergodic risk-sensitive control problem is based on an approximation technique. 
This can be done either by 
discounted risk-sensitive cost models  \cite{cavazos2000vanishing,cavazos2017discounted} (as in Problem 2) or 
by certain discounted risk-sensitive dynamic games, see \cite{cavazos2002alternative,cavazos2011discounted,
hernandez1999existence,hernandez1996risk} for a countable state space case and \cite{di2000infinite,di1999risk,
jaskiewicz2007average,jaskiewicz2007note} for a general state space case. 
This technique leads  via the vanishing discount factor approach to the optimality
equation or to the optimality inequality, 
(when the sign `$=$' in (\ref{eq:ergodic}) is replaced by `$\ge$'). For instance, the existence of 
a solution to the optimality inequality is established in \cite{hernandez1999existence,jaskiewicz2007average} 
where a generalization of the Hardy-Littlewood formula is needed, known as a uniform Tauberian theorem, see
\cite{jaskiewicz2007average} and Proposition 1 in \cite{jaskiewicznowak2014}.
The essential ingredient in this approach is the variational formula for the logarithmic moment-generating function (see
\cite{fleming1997risk,dai1996connections,dembo1998}). 
It should be noted that in contrast to the risk neutral case 
to get a solution to the optimality equation or inequality one needs to assume except ergodicity conditions
that the absolute vale of the risk 
coefficient is sufficiently small. This condition is either imposed explicitly or implicitly, i.e. 
other conditions in fact enforce this requirement, see Example 1 in \cite{jaskiewicz2007average}. 
There is only one exception: the so-called invariant 
models in which the transition probabilities are independent of the state space, 
see \cite{jaskiewicz2007note}.
A further discussion on the conditions when the optimality equation or the optimal inequality 
hold is provided in \cite{cavazos2010optimality}.

The ergodic risk-sensitive control problem is also attacked from different sides. 
 \cite{borkar2002risk} apply an ergodic multiplicative theorem and assume a simple growth condition 
on the one-stage cost function. They establish the optimality equation for a countable state Markov decision chain.
The very recent results for countable state space models have been developed in \cite{biswas2022ergodic,chen2023}.
Finally, an approximation by uniformly ergodic Markov controlled processes 
for a general state space model under minorization condition is studied in \cite{di2007infinite}.
A mutual relationship between the aforementioned works, an extensive discussion of other 
results and a list of further references are given in the 
excellent survey of \cite{biswas2023ergodic}.
Finally, we would like to mention that the nested form of an average risk-sensitive reward is discussed in 
\cite{shen2013risk}.  

Parallel to the theoretical results much effort was put on developing efficient   
algorithms to solve ergodic risk-sensitive control problem. The value iterations are established in 
\cite{bielecki1999value,cavazos2003value} for stationary models 
and in \cite{cavazos2005nonstationary} for non-stationary models.
A Q-learning algorithm is proposed in \cite{borkar2002q} and a version of an actor-critic 
algorithm is considered in \cite{borkar2001sensitivity}. However, these algorithms do not 
incorporate any approximation
of the value function in order to defeat the curse of dimensionality. Such an
approximation in terms of linear combination of a  moderate number of basis functions is developed in 
\cite{basu2008learning}. 
The learning scheme iteratively learns coefficients in the linear combination instead of learning the whole value function. The other tools are applied in \cite{arapostathis2021linear} and
\cite{borkar2017linear} where
equivalent linear and dynamic programs are derived.
The former work deals with minimization of the asymptotic growth rate of the cumulative cost whereas
the latter one uses a  variational representation for asymptotic growth rate of risk-sensitive reward obtained in
\cite{anantharam2017variational}. 
This technique allows to link the average risk-sensitive reward with linear programming
without assuming irreducibility of the Markov chain.  

Except for the average cost/reward criteria defined with the help of an exponential utility function, 
there  are papers that deal with other average risk-sensitive payoff criteria for which 
traditional dynamic programming fails. For example in \cite{cavazos2016characterization} a finite-state irreducible risk-sensitive MDP is considered where the usual exponential utility is replaced by an arbitrary utility function (see also \cite{stettner2023certainty}). The authors prove a connection to the exponential utility criterion. 
\cite{xia2020risk} studies the optimization of the 
mean-variance combined metric assuming that the finite state  
Markov decision chain is {\it ergodic} 
under any stationary policy.  
More precisely, for
$f\in F,$ and an initial  state  $x\in E$ he defines
$${\cal J}^{0}(x,f)=\lim_{n\to\infty} \frac 1n \EE_x^f\left[\sum_{k=0}^{n-1} \left(r(X_k,A_k)-\lambda 
(r(X_k,A_k)-{\cal J}^{av}(x,f))^2\right)\right]$$ 
where $\lambda>0$ is a trade-off parameter and 
$$
{\cal J}^{av}(x,f)=\lim_{n\to\infty} \frac 1n \EE_x^f\left[\sum_{k=0}^{n-1}r(X_k,A_k) \right].$$
Note that ${\cal J}^{av}$ and ${\cal J}^{0}$ are independent of an initial state, because of the ergodicity condition.
The objective is to find a stationary policy $f_*\in F$ which maximizes the associated value, i.e.
$f_*\in\mbox{arg}\max_{f\in F}  {\cal J}^{0}(x,f)$ for all $x\in E.$ Since the optimality equation does not hold, the theory of sensitivity-based optimization is utilized. 
A version of value iteration algorithm is proposed to find an optimal policy.  
The theory of sensitivity-based optimization is also applied in \cite{xia2022risk} to the ergodic Markov 
decision chains when the CVaR measure is used. In this work   \cite{xia2022risk} 
consider the cost functions and aim at the cost functional
$$CVaR_\alpha^f= \lim_{n\to\infty} \frac 1n \sum_{k=0}^{n-1} CVaR^f_\alpha(c_k)$$
where here
$$CVaR^f_\alpha(c_k)=\EE_x^f\left[c(X_k,A_k)|c(X_k,A_k)\ge F^{-1}_{c(X_k,A_k)}(\alpha)\right]$$
and $F^{-1}_{c(X_k,A_k)}(\alpha)$ denotes the upper $\alpha$-quantile  of the random variable $c(X_k,A_k).$
The objective is to find an optimal policy, i.e. $f_*\in F$  such that
$f_*\in\mbox{arg}\min_{f\in F} CVaR_\alpha^f.$ In particular, the authors establish the local optimality equation 
and develop a policy iteration procedure that turns out to be more efficient than solving the bilevel MDP problem
examined among others for risk-sensitive discounted rewards in \cite{bauerle2011avar}. 

At the end let us mention the undiscounted models, i.e models in which the discount factor $\beta=1$
and the time horizon is infinite. 
MDPs with  non-positive payoffs and an entropic risk measure are studied 
in \cite{jaskiewicz2008note}, where   
a non-recursive case is treated (as in Problem 2).
The aim is to show the existence of  an optimal stationary policy and  the convergence of the value iteration algorithm.
In \cite{cavus2014}, on the other hand, a recursive undiscounted cost is defined with the aid of  Markov risk measures.  
For the so-called uniformly risk transient  Markov decision process the optimality equation is established and the existence of an optimal stationary policy.

\section{Applications}\label{sec:appl}

In this section we summarize some applications of the risk-sensitive 
criterion in dynamic, discrete-time optimization problems. 
This is not a complete list but simply a biased selection of examples. 
We start with  the {\em entropic risk measure}. 

\subsection{Entropic Risk Criterion}

One area of applications where the entropic risk criterion is used is {\em financial mathematics and economics}.
In \cite{bielecki1999risk} the authors consider an investment problem in a financial market 
with a factor process given by a Markov chain $(X_t)_{t\in\N}$.  
The evolution of the wealth is defined by
$$ x_{t+1} = x_t \left[ e^r + \pi_t \cdot (Z_{t+1}-e^r 1)\right]$$
where $r$ is a fixed interest rate, $(Z_t)_t$ are the relative price vectors, 
conditionally independent given the states of the Markov chain at time $t$ 
and $t+1$ and $(\pi_t)_t$ are the proportions of wealth invested in the risky assets. 
The aim is to maximize
\begin{equation}\label{eq:potarget}
  \liminf_{T\to\infty} -\frac{2}{\theta} \frac1T \ln \EE_x^\pi \exp{\Big(-\frac{\theta}{2} \ln X_T\Big)} 
\end{equation}
over all investment strategies. Under some irreducibility assumptions an optimal 
investment strategy is stationary and is characterized  
by the  optimality equation given in \eqref{eq:ergodic}.

\cite{stettner1999risk} considers a similar problem which however stems from 
a discretized version of a continuous Black-Scholes model with several factors. 
The optimization criterion is again \eqref{eq:potarget}. Under a uniform ergodicity condition an optimal investment strategy is characterized via the optimality equation. The cases with (proportional) 
and without transaction cost are considered. The model with proportional transaction cost 
and consumption is taken up in \cite{stettner2005discrete}.  
Finally, the assumptions are further relaxed in \cite{pitera2022discrete} for the same optimization criterion.

\cite{bauerle2018stochastic} consider a stochastic optimal growth model 
with nested entropic risk measures. The model is as follows: an agent obtains the output 
$x_t$, which is divided between consumption $a_t$ and investment (saving) $y_t=x_t-a_t$. 
From consumption $a_t$ the agent receives utility $u(a_t)$. 
Investment is used for production with input $y_t$ yielding output 
$$x_{t +1} = f (y_t , \xi_t )$$ 
where $(\xi_t )$ is a sequence of i.i.d. shocks and 
$f$ a production function. 
The criterion of Problem 1 is used for the aggregation of the utilities. 
The value function and an optimal policy are again characterized via the optimality equation. 
Properties of the optimal consumption strategy are also shown. 
The problem is solved explicitly for special utility and production functions. 
The results are extended in \cite{goswami2022regime} to include regime switches.

Other applications in economics touch the problem of precautionary savings, which is one of the most studied issues 
in the theory of choice under uncertainty. For example, \cite{luo2010risk} study  
the consumption-savings behavior of households who have risk-sensitive preferences and 
suffer from limited information-processing capacity (rational inattention). 
The value iteration is as for Problem 1 given by
$$ V(x)= \sup_{c} \Big\{ -\frac12 (c-\bar{c}) -\frac\beta\gamma \ln \EE[\exp(-\gamma V(X_1))]\Big\},$$
where $x$ is the present value of lifetime resources, $c$ is consumption and $\bar{c}$ denotes a bliss point.
The authors  solve the model explicitly and show that rational inattention 
increases precautionary savings 
by interacting with income uncertainty and risk sensitivity.
They show that the model displays
a wide range of observational equivalence properties, 
implying that consumption and savings data cannot distinguish between risk sensitivity, 
robustness, or the discount factor, in any combination. 
 \cite{bommier2019}, on the other hand,  examine non-stationary models of 
precautionary savings with recursive risk-sensitive preferences (as in Problem 1) of the infinitely-lived agents.   Agents are endowed with an exogenous income process $(Z_t).$
The value function in period $t$ is given by the equation
$$V_t(x_t,z^t)=\max_{a_t\in\R} \Big\{\tilde{u} (a_t) -\frac\beta\gamma \ln \EE_t[\exp(-\gamma V_{t+1}(x_{t+1},z^t,Z_{t+1}))]\Big\},$$
where $x_t$ is the wealth at time $t$, $a_t$ is the consumption at time $t$ and $z^t=(z_0,\ldots,z_t)$ is the realized
exogenous income trajectory. Here, $\tilde{u}$ is the one-stage utility of a household.  
It is assumed that the function $(z_0,\ldots, z_t)\to \PP( Z_{t+1}\ge \bar{z}|z_0,\ldots, z_t)$ 
is non-decreasing.   
Moreover,  $a_t>0,$ $x_t+Z_t-y_t=a_t,$ $x_{t+1}=r_{t+1}y_t,$ where $y_t$ is investment and $r_{t+1}$ is 
the deterministic (but time varying) gross interest rate between periods $t$ and $t+1.$  
Additionally, the constraint
$y_t\ge \bar{y}_t(z^t)$ allows to borrow the agent, but no more what she can repay in  the worst scenario. 
The main result announces that the greater risk aversion (the greater absolute values of $\gamma$) implies a higher 
propensity to save at any time. This leads to the conclusion that the greater risk aversion implies greater accumulated wealth or larger precautionary savings. It should be stressed out that this is not the case
when other recursive preferences are considered, for instance, the Epstein-Zin-Weil preferences, see \cite{epstein2013substitution,weil1990nonexpected} 
or  the preferences developed in \cite{weil1993precautionary}. 
The reader is referred to the numerical results obtained in \cite{bommier2019}.    

It is worth mentioning that Pareto optimal consumption allocations is studied by \cite{anderson2005dynamics},
who also assumes that the agents have recursive risk-sensitive preferences defined by an exponential utility function.

Nested entropic risk measures are used in actuarial theory as well. In this matter the reader is referred 
to the works of  \cite{bauerle2015risk, bauerle2017optimal}. 
In the latter paper, within the recursive preference framework they determine 
the optimal dividend strategy for an insurance company and 
derive a policy improvement algorithm.

The next prominent applications can be found in the {\em operations research} area. 
The paper of  \cite{bsobel1992inventory} is one of the first paper that uses the exponential
utility function to the multiperiod news vendor model. 
The authors  minimize the risk-sensitive discounted cost, i.e. 
as in Problem 2. It is shown that the base-stock policy is optimal and depends on the length of a time horizon, 
discount factor and risk parameter. For the infinite time horizon an optimal policy is ultimately stationary.
Their considerations are extended to models with dependent demands in \cite{choi2011} where
an asymptotic behavior of the
solution when the degree of risk aversion coefficient converges
to zero or infinity is analyzed. 
Another interesting issue from the area of {\em revenue management} can be found in \cite{barz2007risk}.
The approach is explained in the setting of optimal airline ticket booking where 
the airline has to decide whether or not to accept a request for a certain fare 
given the remaining capacity. The target function is the one from Problem 2. 
The optimal strategy is computed and compared to the risk-neutral setting. 
Further applications to revenue management with different risk-averse target 
functions can be found in \cite{schlosser2015stochastic,schlosser2016stochastic}. 
A survey of  risk-sensitive and robust revenue management problems the reader may find in \cite{gonsch2017survey}, 
where among other issues the  capacity control and dynamic pricing are considered. 
Finally, \cite{denardo2007risk} consider the multiarmed bandit problem with an exponential 
utility and criterion as in Problem 2. They show the optimality of some kind of index policy 
using analytical arguments.

Applications in {\em computer science and engineering} are as follows. 
One of the first papers is \cite{koenig1994risk}. The authors
discuss goal reaching problems (e.g.\ for robots) under risk-sensitive criteria. 
They obtain the following optimality equation (there is no discounting):
$$ V(x)=\inf_{a} \left\{ \sum_{y\in E\setminus G} q(y|x,a) e^{\gamma c(x,a,y)} V(y) 
+ \sum_{y\in  G} q(y|x,a) e^{\gamma (c(x,a,y)+r(y))} \right\} $$
where $G\subset E$ are the goal states, $c(x,a,y)$ is the cost of executing action $a$ in state $x$ 
and proceeding state $y$ and $r$ is the terminal reward function. Solution algorithms, 
in particular under change of measure are discussed and some block world problems 
are considered. 
In \cite{medina2012risk,befekadu2015risk} the authors consider a finite time horizon 
linear-quadratic problem with target function like in Problem 2 with an exponential utility. 
In \cite{medina2012risk} the setting is  to optimize a human-robot interaction 
such that the physically coupled human-robot follows a desired trajectory. 
\cite{befekadu2015risk} study the impact
of cyber-attacks in control systems with partial observation. 
Further, \cite{mazouchi2022automating} investigate risk-averse preview-based 
Q-learning planner  for navigation of autonomous vehicles on a multi-lane road. 
The criterion is that of Problem 2 with an exponential utility function. 

\subsection{CVaR Risk Criterion}

Another popular optimization criterion is the CVaR. 

We start with some examples from {\em operations research and engineering}.
\cite{gonsch2018optimizing} consider  dynamic pricing with a risk-averse seller maximizing 
the CVaR over the selling horizon. The aim is to dynamically adjust the price during the 
selling horizon in order to  sell a fixed capacity of a perishable product  
where demand is stochastic such that the total expected/risk averse revenue is maximized. 
As optimization criterion they use the CVaR of the cumulated revenue. More precisely, 
they consider the setting of Section \ref{sec:outside} with a finite time horizon and CVaR, i.e.
$$ \max_{\pi\in\Pi} CVaR_\alpha\Big(\sum_{k=1}^N A_k 1_{[Y_k\ge A_k]} \Big) =: V_N(x)$$
where $A_k$ is the price offered at time $k$ by the firm. The state $x$ is the remaining good 
and $Y_k$ are i.i.d. continuous random variables which represent the willingness 
to pay of a potential customer arriving in period $k.$
The authors use recursive algorithms to solve the problem, 
based on specific properties of the CVaR 
given by $V_0(x,\alpha)=0$ for $x\ge 0$ and
\begin{eqnarray*}
 \lefteqn{ V_t(x,\alpha) =}\\&& \max_a 
 CVaR_\alpha \Big\{1_{[Y_t\ge a]} \left(a+ V_{t-1}(x-1, 
 \alpha z_{t-1,x-1})\right) + 1_{[Y_t< a]}  
V_{t-1}(x, \alpha z_{t-1,x}) \Big\}  
\end{eqnarray*}
where $z_{t-1,x-1}$ are certain constants arising from CVaR minimization. A nested formulation with CVaR is considered in \cite{schur2019time}.

\cite{wozabal2020optimal} consider multi-stage stochastic programming 
approaches to optimize the bidding strategy of a virtual power plant operating 
on the Spanish spot market for electricity. They consider different setups among others 
a nested CVaR approach.

 \cite{maceira2014application}  deal with hydrothermal generation planning in Brazil. 
The aim is to  optimize the system operation, taking into account the expected value of thermal 
generation and possible load curtailment costs over a given set of inflow scenarios to the reservoirs 
in the future. Risk aversion is crucial here  to avoid unacceptable amounts of 
load curtailment in critical inflow scenarios. The authors use nested CVaR and  
dual stochastic dynamic programming to solve the problem.

The PhD thesis of  \cite{ott2010markov} treats several problems of surveillance of 
critical infrastructures treated as stochastic dynamic optimization problems. 
The author uses CVaR as criterion in the total discounted cost problems and average cost problems.

 \cite{jiang2016practicality} investigate a dynamic decision problem faced by the manager 
of an electric vehicle  charging station, who aims to satisfy the charging demand of the customer 
while minimizing cost. Since the total time needed to charge the electric vehicle up to capacity 
is often less than the amount of time that the customer is away, 
there are opportunities to exploit electricity spot price variations. 
The authors formulate this problem as a combination of nested CVaR and 
expectation over a finite time horizon. They identify structural properties 
of the optimal policy and  propose an approximation algorithm based on regression 
and polynomial optimization to solve the problem.

 \cite{zhang2016decomposition} consider five decompositions of nested CVaR application in   multistage  stochastic  linear  programming. They apply the proposed formulations 
to a water management problem in the area of the southeastern portion of Tucson, AZ 
to best use the limited water resources available to that region.

Finally, \cite{ahmed2007} solve a multiperiod inventory model with nested approach
of coherent risk measures. 
For a finite time horizon they prove  that the optimal policy has a similar 
structure as that of the expected value problem. Moreover, an analyis of monotonicity properties of the
optimal order quantity with respect to the degree of risk aversion for certain risk measures 
like CVaR is conducted. 

Applications in {\em financial mathematics and economics} are as follows: \cite{staino2020nested} treat  portfolio optimization problems with nested 
CVaR when asset log returns are stage-wise dependent by a single-factor. 
Using a cubic spline interpolation the authors numerically solve the problem 
with a finite time horizon by backward recursion. A dynamic mean-risk problem, where the risk constraint is
given by the CVaR is considered in \cite{bauerle2009dynamic}. 
The financial market is a binomial model which allows for explicit solutions. 
Since the  problem is solved via a Lagrange function, the CVaR appears in the optimization criterion. 
It is applied to the cumulated gain/loss and the problem is solved by recursion explicitly.

An application in {\em biology} is given in
\cite{bushaj2022risk} where the authors  
apply a mean-CVaR  multistage, stochastic mixed-integer programming model to optimize 
a manager’s decisions about the surveillance and control of a non-native forest insect, 
the emerald ash borer. 

As mentioned before, this is just a selection of applications. Further examples can be found in the literature.


\begin{appendices}

\section{Proof of Theorem \ref{theo:prob1}}\label{secA1}

First we show the statements under assumption (W). 
Let $v\in U(E)$ and define 
$$ Lv(x) =  \sup_{a\in D(x)} \Big\{ r(x,a) + \beta S_u^{(x,a)}(v(X_1)) \Big\}$$
where $X_1$ has distribution $q(\cdot|x,a)$. We first prove that $L : U(E) \to U(E).$ 
Note that by  (P1) and (P4) we get for every $x\in E$
 \begin{align*}
            Lv(x)&= \sup_{a\in D(x)} \Big\{ r(x,a) + \beta S_u^{(x,a)}(v(X_1)) \Big\}\ge 0+\sup_{a\in D(x)}  \beta S_u^{(x,a)}(0)\ge 0.
        \end{align*}
On the other hand, we have again by (P1) and (P4) that
 \begin{align*}
            Lv(x)
            &\le d+\sup_{a\in D(x)} \beta S_u^{(x,a)}(\|v\|)= d+\beta \|v\|.
        \end{align*}
Now we show that $Lv$ is upper semicontinuous. For this purpose we prove for $v\in U(E)$ that
\begin{align}\label{eq:intusc}
    (x,a,\eta) \to r(x,a) + \beta \eta + \beta \int u(v(y)-\eta) q(dy|x,a)
\end{align}   
is upper semicontinuous.  Clearly, 
$(x,a,\eta) \to r(x,a) + \beta \eta$ is upper semicontinuous.
For the second part
assume that $(x_n,a_n,\eta_n)$ is a 
sequence which converges to $ (x_0,a_0,\eta_0)$ as 
$n\to\infty$ where $x_n\in E,$ $a_n\in D(x_n),$ $\eta_n
\in\R$  for $n\in\N_0.$ Set
$\phi_n(y):=u(v(y)-\eta_n)$ for $n\in\N.$ Since $u$ is 
continuous and non-decreasing, $\phi_n$ are upper semicontinuous. Making use of the Fatou lemma for weakly convergent measures (see Lemma 3.6 in \cite{jota}) we get that
$$
\limsup_{n\to\infty} \int \phi_n(y) q(dy|x_n,a_n)\le
\int \phi^*(y)q(dy|x_0,a_0)  
$$
with $\phi^*(x)=\sup\{\limsup_{n\to\infty} \phi_n(y_n): y_n\to x\}.$ The supremum is taken over all sequences $(y_n)$
converging to $x$. In our case, for any $y_n\to x$
$$
\limsup_{n\to\infty} \phi_n(y_n)=
\limsup_{n\to\infty} u(v(y_n)-\eta_n)\le u(v(x)-\eta_0).
$$
Hence, $\phi^*(x)=u(v(x)-\eta_0).$ This proves that the function in \eqref{eq:intusc} is upper semicontinuous.

Next we conclude by Proposition 2.1 in \cite{ben2007old} 
that the supremum over all $\eta\in\R$ in the definition of the Optimized 
Certainty Equivalent can be restricted to the compact set,
for example $[0,\|v\|].$ This is the support of the random variable $v(X_1).$ Hence, by Proposition 2.4.3 in \cite{bauerle2011markov} the function 
 \begin{align*}
            Lv(x)
            &= \sup_{a\in D(x)} \sup_{\eta\in [0,\|v\|]}  \Big\{ r(x,a) + 
						\beta \eta + \beta \int u(v(y)-\eta) q(dy|x,a) \Big\}.
        \end{align*}
is upper semicontinuous. 

Finally we prove that $L$ is contracting. Let $v_1,v_2\in U(E).$ Then due (P1) and (P2) 
and  $v_1 \le v_2 + \|v_1-v_2\|$,  we obtain:
\begin{align*}
    Lv_1(x)-Lv_2(x) &\le \beta \sup_{a\in D(x)} \left(S_u^{(x,a)}(v_1(X_1))-  S_u^{(x,a)}(v_2(X_1)) \right)\\
    &\le  \beta \sup_{a\in D(x)} \left(S_u^{(x,a)}(\|v_1-v_2\| + v_2(X_1))-  S_u^{(x,a)}(v_2(X_1))\right)\\
	&	=\beta \|v_1-v_2\|.
\end{align*}
Interchanging the roles of $v_1$ and $v_2$ yields $\| Lv_1-
Lv_2\|\le \beta \|v_1-v_2\|. $ 
Finally since $U(E)$ equipped with the supremum norm is 
complete, the Banach fixed point theorem implies that there 
exists $V\in U(E)$ such that $V=LV.$

It remains to show that $V$ is the value function. Observe that for all $(x,a)\in D$  we immediately have
$$ V(x) \ge r(x,a)+ \beta S_u^{(x,a)}(v(X_1)).$$
Let $(\pi_k)_{k\in\N_0}\in\Pi$ be any policy. Then for all $k=1,\ldots ,N$ we obtain
$ V(x_k) \ge L_{\pi_k} V(h_k).$ Making use of this inequality  by iteration  we infer that
$$ V(x)\ge (L_{\pi_0} \circ \ldots \circ L_{\pi_N}) V(x)\ge (L_{\pi_0} 
\circ \ldots \circ L_{\pi_N}) \mathbf{0}(x) =J_{N+1}(x,\pi). $$
Letting $N\to\infty$ implies $V(x)\ge J(x,\pi)$ 
for all policies $\pi\in\Pi$ which in turn gives 
\begin{equation}\label{oo}
V(x) \ge \sup_{\pi\in\Pi} J(x,\pi)\quad\mbox{for every}\quad x\in E.
\end{equation}
For the reverse inequality by Proposition 2.4.3 in \cite{bauerle2011markov}  it follows that firstly
the function
$$(x,a)\to \sup_{\eta\in[0,\|V\|]}\left\{r(x,a)+\beta \eta+
\beta\int u(V(y)-\eta)q(dy|x,a)\right\}$$
is upper semicontinuous and secondly,
there exists  $f^*\in F$ such that $V = L_{f^*}^{(N)}V.$ 
Thus, again by iteration we have $V = L_{f^*}^{(N)}V$ 
where $L_{f^*}^{(N)}$ denotes the composition of $L_{f^*}$ with itself $N$ times.
Hence, putting $r(x,f^*(x))=r_{f^*}(x)$ we get
\begin{align*}
    V(x) &\le  L_{f^*}^{(N-1)}\Big( r_{f^*}+\beta \|V\|\Big)(x)
    \\
    &=  L_{f^*}^{(N-2)}
		\Big( r_{f^*}+\beta S^{(\cdot,f^*(\cdot))}_u( r_{f^*}(X_1))+\beta^2  \|V\|\Big)(x)
		\\
    &\le \ldots\le 
    J_N(x,f^*) +\beta^N \|V\|.
\end{align*}
Letting $N\to\infty$ yields that $V(x)\le J(x,f^*)$ for every $x\in E.$ This fact and (\ref{oo}) finish the proof. 

Assume now that (S) holds. It suffices to show that 
$L:B(E)\to B(E).$ Let $v\in B(E).$ 
Assume that $(a_n,\eta_n)\to(a_0,\eta_0)$ as $n\to\infty$
for $a_n\in D(x)$ and $\eta_n\in\R.$ Then, by condition (S)
and Proposition 18 
on p. 270 in \cite{royden} we have that
$$\int u(v(y)-\eta_n)q(dy|x,a_n) \to \int u(v(y)-\eta_0)q(dy|x,a_0) \quad\mbox{as } n\to\infty.
$$
Hence, the function
$$ (a,\eta)\to \left\{r(x,a)+\beta \eta+\beta\int 
u(v(y)-\eta)q(dy|x,a)\right\}
$$
is upper semicontinuous for each $x\in E$. Again the measurable selection theorem (see Theorem A.2.4 in \cite{bauerle2011markov}) and the fact that 
 by Proposition 2.1 in \cite{ben2007old} 
 the supremum over all $\eta\in\R$ in  $S_u^{(x,a)}$ can be replaced by the supremum over the set  $[0,\|v\|],$ imply
 that $Lv\in B(E).$ Now the remaining part proceeds along the same lines with obvious changes, i.e. the fixed point of $L$ is found in $B(E).$
 
\section{Proof of Theorem \ref{theo:prob2}}\label{secA2}%

The proof of part a) is essentially Theorem 3 in \cite{bauerle2014more}. 
The only difference is that we have a maximization problem here instead of a minimization problem.

For part b) note again that $R_\beta^\infty $ is bounded and thus the 
maximization over $\eta$ in the definition of $S_u$ can be restricted to 
a compact set by Proposition 2.1 in \cite{ben2007old}. In other words, 
we have to solve in the second step for large $K>0$ 
$$  \sup_{\eta\in [-K,K]}  \Big\{  \eta + V_\infty(x,-\eta,1)\Big\}. $$
But from part a) we know  that $V_\infty$ is continuous in $\eta$ 
which implies the existence of an $\eta^*$ with
$$  \sup_{\eta\in [-K,K]}  \Big\{  \eta + V_\infty(x,-\eta,1)\Big\} = \eta^* + V_\infty(x,-\eta^*,1)$$
and thus the statement.
\end{appendices}

\bibliography{literature_RS}
\end{document}